\documentclass[aps,prb,preprint]{revtex4}
\usepackage{amsmath}
\usepackage{graphicx}

\input{tcilatex}

\begin{document}

\title{Field-tuned quantum tunneling in a supramolecule dimer [Mn$_{4}$]$%
_{2} $}
\author{Yuanchang Su$^{1,2}$ and Ruibao Tao$^{3,2}$
\footnote {To whom correspondence should be addressed. Email: rbtao@fudan.edu.cn}%
}
\affiliation{{\footnotesize 1.State Key Laboratory of Applied Surface Laboratory, Fudan
University, Shanghai 200433, China}\\
{\footnotesize {2.Department of Physics, Fudan University, Shanghai 200433,
China}}\\
{\footnotesize 3. Chinese Center of Advanced Science and Technology (World
Laboratory) , P. O. Box 8730 Beijing 100080, China}}

\begin{abstract}
Field-tuned quantum tunneling in two single-molecule magnets
coupled antiferromagnetically and formed a supramolecule dimer is
studied. We obtain step-like magnetization curves by means of the
numerically exact solution of the time-dependent Schr\H{o}dinger
equation. The steps in magnetization curves show the phenomenon of
quantum resonant tunneling quantitatively. The effects of the
sweeping rate of applied field is discussed. These results
obtained from quantum dynamical evolution well agree with the
recent experiment[W.Wernsdorfer et al. Nature
416(2002)406].\newline

\medskip APCS Number: 75.50.Xx,75.45.+j,76.20.+q
\end{abstract}

\maketitle

The macroscopic quantum phenomena in molecular magnets has become a very
attractive researching field. Many properties of these nanometer-sized
magnetic particles and clusters, such as Mn$_{12}$($s=10$), Fe$_{8}$($s=10$)
and Mn$_{4}$($s=\frac{9}{2}$) systems, have been well studied$^{[1]-[6]}$
both experimentally and theoretically. Theoretically, studying the
phenomenon of quantum resonant tunneling of these molecular magnets could be
based on Landau-Zener (LZ) transitions$^{[7],[8]}$, or based on numerically
the solution of the time-dependent Schr\H{o}dinger equation$^{[9],[10]}$. In
Landau-Zener model, the magnetization curves could be obtained in a static
and approximate way. Recently a supermolecular dimer [Mn$_{4}$]$_{2}$ is
reported to be synthesized successfully by Werndorfer et al.$^{[11]}$. In
this kind of supermolecular dimer, two single-molecule magnets Mn$_{4}$
antiferromagnetic coupled each other, which results in its quantum behavior
quite different from two individual Mn$_{4}$ molecule without coupling. In
this paper, we calculate magnetization curves of a supermolecular dimer [Mn$%
_{4}$]$_{2},$a single-molecule magnets, by numerically exact solution of the
time-dependent Schr\H{o}dinger equation.

Following Wernsdorfer et al., the model Hamiltonian of the supermolecular
dimer [Mn$_{4}$]$_{2}$ is
\begin{equation}
H=H_{1}+H_{2}+JS_{1}\cdot S_{2},  \label{1}
\end{equation}%
where $J$ is the weak antiferromagnetic supercharging coupling constant. H$%
_{1}$ and H$_{2}$ are Hamiltonian for two individual Mn$_{4}$ molecules in
the supermolecular dimer. It is known that the model Hamiltonian of an
individual Mn$_{4}$ molecule is
\begin{equation}
H_{i}=-DS_{zi}^{2}+E(S_{xi}^{2}-S_{yi}^{2})-g\mu _{B}S_{zi}h_{z}(t),i=1,2,
\label{2}
\end{equation}%
where $D$ and $E$ are the axial anisotropic constants. $h_{z}(t)$\ is the
applied sweeping field along easy axis. We can easily obtain the energy
eigenvalues of whole Hamiltonian $H$\ (Figure 1).\ In experiment the
sweeping rate of $h_{z}(t)$ is very slow, so we can simulate it as a
step-increased field, which means the $h_{z}$\ increases a value $\Delta
h_{z}$\ every $\tau $\ time step and keeps constant during the $\tau $\ time
intervals. Note that we can not use a sweeping rate as slow as experiment
due to the limitation of our computing time. However, we can obtain the key
macroscopic quantum phenomena in our calculation with relatively high
sweeping rate. In this paper, we select $D=0.72k,J=0.1k,$$^{[11]}$and $%
E=0.0317k$$^{[6]}$. Dynamic evolution follows time-dependent schr\H{o}dinger
equation and can be calculated by
\begin{equation}
\left\vert \Psi \left( t\right) \right\rangle =\left\vert \Psi \left(
t_{0}+n\tau \right) \right\rangle =\exp [-iH(t_{0}+(n-1)\tau )\cdot \tau
]\left\vert \Psi \left( t_{0}+(n-1)\tau \right) \right\rangle ,  \label{3}
\end{equation}%
Meanwhile, $\left\vert \Psi \left( t\right) \right\rangle $ can be expanded
as
\begin{equation}
\left\vert \Psi \left( t\right) \right\rangle =\overset{S}{\underset{m_{1}=-s%
}{\sum }}\overset{S}{\underset{m_{2}=-s}{\sum }}a_{m_{1},m_{2}}\left(
t\right) \left\vert m_{1},m_{2}\right\rangle ,S=\frac{9}{2},  \label{4}
\end{equation}%
where $\left\vert m_{1},m_{2}\right\rangle $ are the eigenstates of $H_{0}$
that is
\begin{equation}
H_{0}=-DS_{z1}^{2}-DS_{z1}^{2}+JS_{z1}\cdot S_{z2}-g\mu
_{B}(S_{z1}+S_{z2})h_{z}(t),  \label{5}
\end{equation}%
We assume the initial states to be at $\left\vert -\frac{9}{2},-\frac{9}{2}%
\right\rangle $, and the whole evolution process can be obtained by equation
(4) step by step.\ In Ref.$^{[11]}$, Wernsdorfer report five points (Figure
4 of Ref.) of resonant tunneling that result in the steps in hysteresis
loops. They considered the first point is caused by the resonant transition
from\ $\left\vert -\frac{9}{2},-\frac{9}{2}\right\rangle $\ to$\left\vert -%
\frac{9}{2},\frac{9}{2}\right\rangle $, and the fourth point is caused by
those from $\left\vert -\frac{9}{2},-\frac{9}{2}\right\rangle $\ to$%
\left\vert -\frac{9}{2},\frac{5}{2}\right\rangle $. However, under the model
Hamiltonian of Equation(1) and Equation(2), the transitions of these points
are quenched for a half integer spin due to the parity symmetry$^{[12],[13]}$%
. Therefore, there must be some kind of transverse field components resulted
from the influence of the environmental degrees$^{[6]}$, such as hyperfine
and dipolar couplings, and it can be approximated a Gaussian distribution
with a width $\sigma =0.035T$ for such additional transverse environmental
field. In this paper, we simply assume it to be a constant and $h_{x}=0.01T$
along x axis, but do not lose the essential physics, the macroscopic quantum
phenomena.

The magnetization along the $z$ axis can be simply defined by $%
M=<S_{z1}+S_{z2}>$. In Figure 2, we plot the magnetization curve responding
to a time-dependent applied field with a constant transverse field $%
h_{x}=0.01T$. There are three steps in the magnetization curve. In order to
know the details of state transitions, the states ($\left\vert \Psi \left(
t\right) \right\rangle $) near to two sides of resonant points are recorded
in our simulation and they are shown in Table 1, where the occupied
probabilities ($\left\vert a_{m_{1},m_{2}}\left( t\right) \right\vert ^{2}$%
)\ are neglected to zeros if they less than $10^{-3}$. Therefore, we can get
clear information of the process of evolution and transition. Figure 1 and
Table 1 show that the first step occurs at $h_{z}=0.198T$\ from $\left\vert -%
\frac{9}{2},-\frac{9}{2}\right\rangle $\ to$\left\vert -\frac{9}{2},\frac{7}{%
2}\right\rangle $ (and$\left\vert \frac{7}{2},-\frac{9}{2}\right\rangle $),
and the second step occurs at $h_{z}=0.731T$\ from $\left\vert -\frac{9}{2},-%
\frac{9}{2}\right\rangle $\ to$\left\vert -\frac{9}{2},\frac{5}{2}%
\right\rangle $ (and$\left\vert \frac{5}{2},-\frac{9}{2}\right\rangle $).
These two resonant points fit well to experimental results (i.e. the point 2
and the point 4 of Fig.4 in Ref.$^{[11]}$). The third step occurs at $%
h_{z}=0.812T$\ from $\left\vert -\frac{9}{2},\frac{5}{2}\right\rangle $\ (and%
$\left\vert \frac{5}{2},-\frac{9}{2}\right\rangle $) to$\left\vert \frac{7}{2%
},\frac{5}{2}\right\rangle $ (or$\left\vert \frac{5}{2},\frac{7}{2}%
\right\rangle $). There is no step at $h_{z}\approx -0.33T$ in our
magnetization curve, but the experiment reports a point of resonant
tunneling (the point 1 in Fig.4 of Ref.$^{[11]}$). The reason is that we
have used a too fast sweep rate in our simulation. We will interpret it more
detail late. In our magnetization curve, since the step at $h_{z}\approx
-0.33T$ from $\left\vert -\frac{9}{2},-\frac{9}{2}\right\rangle $\ to$%
\left\vert -\frac{9}{2},\frac{9}{2}\right\rangle $ does not occur, therefore
the step at $h_{z}=0.87T$ from $\left\vert -\frac{9}{2},\frac{9}{2}%
\right\rangle $\ to$\left\vert \frac{7}{2},\frac{9}{2}\right\rangle $ (point
4 in the Fig.4 of Ref.$^{[11]}$) can not occur naturally.

In order to interpret why there is no step at $h_{z}=-0.33T$ in our
magnetization curve, we firstly consider a utmost-slow process. At any $t$
value, the eigenstates and eigenvalues (Figure 1) of whole Hamiltonian $H(t)$%
\ can be calculated by
\begin{equation}
H(t)\left\vert \Phi (E)\right\rangle =E\left\vert \Phi (E)\right\rangle ,
\label{6}
\end{equation}%
Note that no matter how weak it is, the system always interact with
environment which cause dissipation. Therefore, in a very very slow process,
we can assume that the state $\left\vert \Psi (t)\right\rangle $\ of system
evaluating from a initial state $\left\vert \Psi (0)\right\rangle $ will
always relax to the ground state $\left\vert \Phi (E_{\min })\right\rangle $%
\ of $H(t)$.\ Figure 3 shows the magnetization curve of this utmost-slow
process. There are two steps at $h_{z}\approx -0.3363T$ and $h_{z}\approx
0.3363T$ in the curve. It means that the point of resonant tunneling at $%
h_{z}\approx -0.3363T$ (point 1 of Fig.4 of Ref.$^{[11]}$) occurs when the
sweeping rate of applied field is very slow. In figure 2, the sweeping rate
of applied field in our calculation is $c=\frac{\Delta h_{z}}{\tau }=\frac{%
10^{-5}}{10^{-8}}=1000$ $Tesla/s$, which is much more larger than the ones
in experiment ($0.140$ $Tesla/s$, $0.035$ $Tesla/s$ and $0.004Tesla/s$)$%
^{[11]}$. Due to the limitation of our computer time, we can not do the
calculation for a sweeping rate of applied field as slow as the one in
experiment. We now try to simulate the magnetization process (Figure 4) only
in a very sharp range of $h_{z}$ with sweeping rates as slow as the ones
used in experiment. In our figure 4, (a) is calculated over a range of $%
h_{z} $ from $-0.3364$ $Tesla$ to $-0.3362$ $Tesla$ with parameters $\tau
=10^{-8}s $\ and $\Delta h_{z}=10^{-9}Tesla$ (the sweeping rate $\sim 0.10$ $%
Tesla/s$); (b) is calculated over a range of $h_{z}$ from $-0.336295$ $Tesla$
to $-0.336275$ $Tesla$ with parameters $\tau =10^{-8}s$\ and $\Delta
h_{z}=10^{-10}$ $Tesla$ ( the sweeping rate $\sim 0.01Tesla/s$); (c) is the
combination of (a) and (b). A very clear step occurs at $h_{z}\approx
-0.336283T$ point, and it shows that the slower sweeping rate induces the
higher step occurred. The recorded states (Table 2) at transition point $%
h_{z}\approx -0.3363$ $Tesla$ show that the resonant tunneling is from $%
\left\vert -\frac{9}{2},-\frac{9}{2}\right\rangle $\ to$\left\vert -\frac{9}{%
2},\frac{9}{2}\right\rangle $ (and$\left\vert \frac{9}{2},-\frac{9}{2}%
\right\rangle $). All these results fit well with the results of experiment$%
^{[11]}$. Therefore, it clear show that the reason for no step at point
about $h_{z}\approx -0.3363Tesla$ is from too fast sweeping rate of the
applied field in theoretical simulation. There are some small oscillations
in the magnetization curve. It is caused from quantum fluctuations.

In conclusion, We have studied the phenomenon of quantum resonant tunneling
in a supermolecular dimer [Mn$_{4}$]$_{2}$ of single-molecule magnets by
numerically exact solution of the time-dependent Schr\H{o}dinger equation.
We obtain step-like magnetization curves which demonstrate quantum tunneling
quantitatively. We have calculated and discussed the affect to steps caused
by different sweeping rate of applied field. It shows that some steps can
not occur at some resonant points when the sweeping rate is too fast, but
they could appear when the sweeping rate becomes enough slow. At a very
narrow region near resonant point, we slow down the sweeping rate of applied
field, some quantum resonant tunnelling appeared in experiment can appear.
Meanwhile, theoretical calculation show that more slow rate induces more
higher transition step. The results of our calculation fit very well with
the experiment$^{[11]}$. Note that since we do not take into account the
effects of dissipation caused by environment, the magnetization curves we
obtain can not reach a reversal saturation value even if the applied field
increase to infinitive value. Therefore, if we want to calculate a whole
hysteresis loop, a proper mechanism of dissipation should be taken into
account.

This work is supported by National Natural Science Foundation of China.

\newpage

{\LARGE caption}

{\large Table1:} Occupied probabilities $\left| a_{m_{1},m_{2}}\left(
t\right) \right| ^{2}$ at spin states $\left| m_{1},m_{2}\right\rangle $
(Equation (5)) at some points of the evolution process (Figure 2). The
values are neglected to zero if they are smaller than $0.001$.

\begin{tabular}{|l|l|l|l|l|l|l|l|l|l|}
\hline
$h_{z}(Tesla)$ & -0.34 & -0.32 & 0.19 & 0.21 & 0.72 & 0.74 & 0.80 & 0.82 &
0.9 \\ \hline
$\left\vert -\frac{9}{2},-\frac{9}{2}\right\rangle $ & 0.9986 & 0.9987 &
0.9987 & 0.9861 & 0.9855 & 0.9719 & 0.9718 & 0.9718 & 0.9717 \\ \hline
$\left\vert -\frac{9}{2},\frac{7}{2}\right\rangle $ & 0 & 0 & 0 & 0.0062 &
0.0061 & 0.0060 & 0.0062 & 0.0061 & 0.0060 \\ \hline
$\left\vert \frac{7}{2},-\frac{9}{2}\right\rangle $ & 0 & 0 & 0 & 0.0062 &
0.0061 & 0.0060 & 0.0062 & 0.0061 & 0.0060 \\ \hline
$\left\vert -\frac{9}{2},\frac{5}{2}\right\rangle $ & 0 & 0 & 0 & 0 & 0 &
0.0068 & 0.0067 & 0.0053 & 0.0053 \\ \hline
$\left\vert \frac{5}{2},-\frac{9}{2}\right\rangle $ & 0 & 0 & 0 & 0 & 0 &
0.0068 & 0.0067 & 0.0053 & 0.0053 \\ \hline
$\left\vert \frac{7}{2},\frac{5}{2}\right\rangle $ & 0 & 0 & 0 & 0 & 0 & 0 &
0 & 0.0013 & 0.0013 \\ \hline
$\left\vert \frac{5}{2},\frac{7}{2}\right\rangle $ & 0 & 0 & 0 & 0 & 0 & 0 &
0 & 0.0013 & 0.0013 \\ \hline
$Total$ & 0.9986 & 0.9987 & 0.9987 & 0.9985 & 0.9977 & 0.9975 & 0.9977 &
0.9973 & 0.9970 \\ \hline
\end{tabular}

\newpage

{\large Table2:} Occupied probabilities $\left|
a_{m_{1},m_{2}}\left( t\right) \right| ^{2}$ at spin states
$\left| m_{1},m_{2}\right\rangle $ (Equation (5)) at some points
of the evolution process (Figure 4). The values are neglected to
zero if they are smaller than $0.0001$.

\begin{tabular}{|l|l|l|l|l|}
\hline
Figure 4 & \multicolumn{2}{|c|}{(a)} & \multicolumn{2}{|c|}{(b)} \\ \hline
$h_{z}(Tesla)$ & -0.3364 & -0.3362 & -0.336295 & -0.336275 \\ \hline
$\left\vert -\frac{9}{2},-\frac{9}{2}\right\rangle $ & 1 & 0.9966 & 1 &
0.9679 \\ \hline
$\left\vert -\frac{9}{2},\frac{9}{2}\right\rangle $ & 0 & 0.0016 & 0 & 0.0160
\\ \hline
$\left\vert \frac{9}{2},-\frac{9}{2}\right\rangle $ & 0 & 0.0016 & 0 & 0.0160
\\ \hline
$Total$ & 1 & 0.9998 & 1 & 0.9999 \\ \hline
\end{tabular}

\newpage

\begin{figure}[p]
\center{\includegraphics[width=16cm]{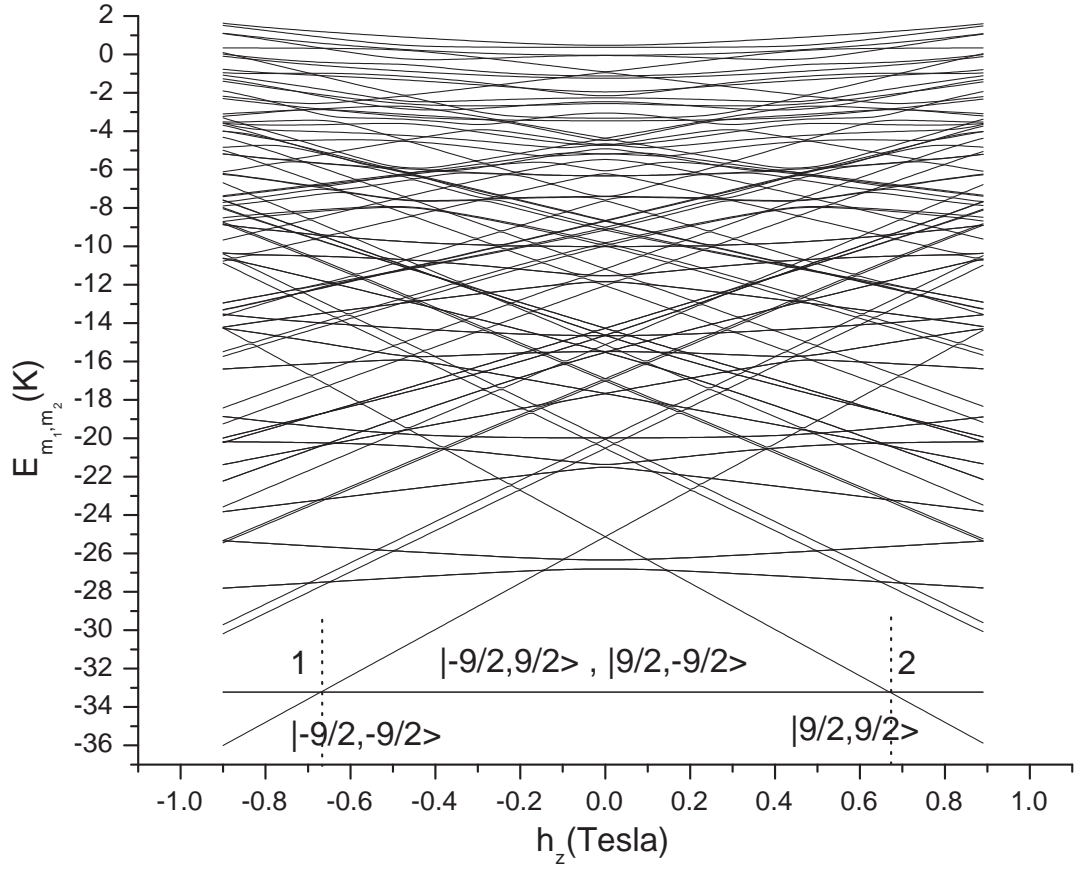}}
\caption{The 100 spin state energies of the model Hamiltonian(Equation (1))
as a function of longitude applied field. A weak transverse field $%
h_{x}=0.01Tesla$\ is take into account.}
\label{fig:epsart}
\end{figure}

\begin{figure}[tbp]
\center{\includegraphics[width=16cm]{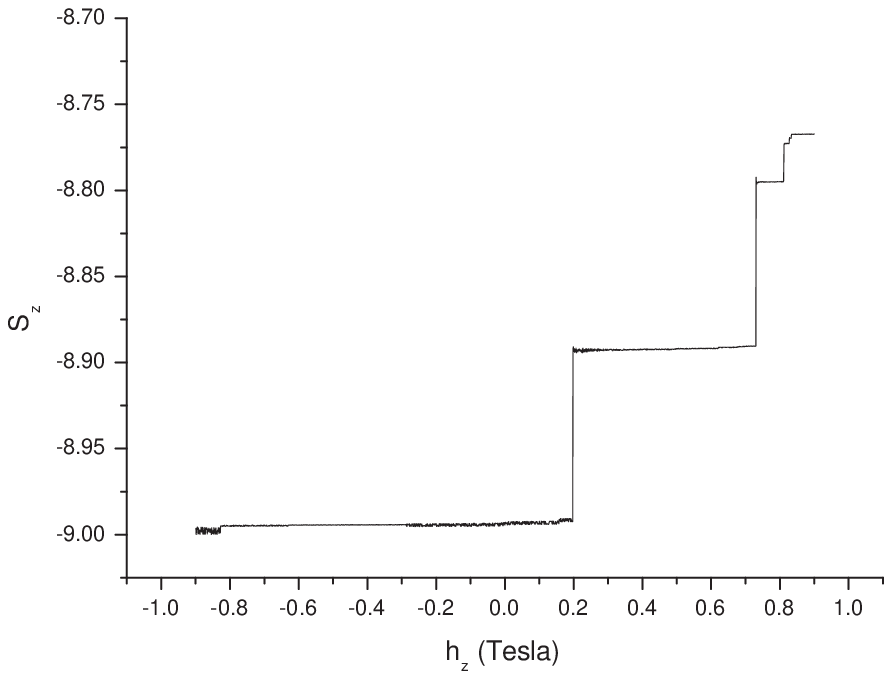}}
\caption{Magnetization relaxation of ground state (Magnetization curve
reponse to a sweeping field. The sweeping rate is $c=\frac{\Delta h_{z}}{%
\protect\tau }=\frac{10^{-5}}{10^{-8}}=1000Tesla/s$.}
\label{fig:epsart}
\end{figure}

\begin{figure}[tbp]
\center{\includegraphics[width=16cm]{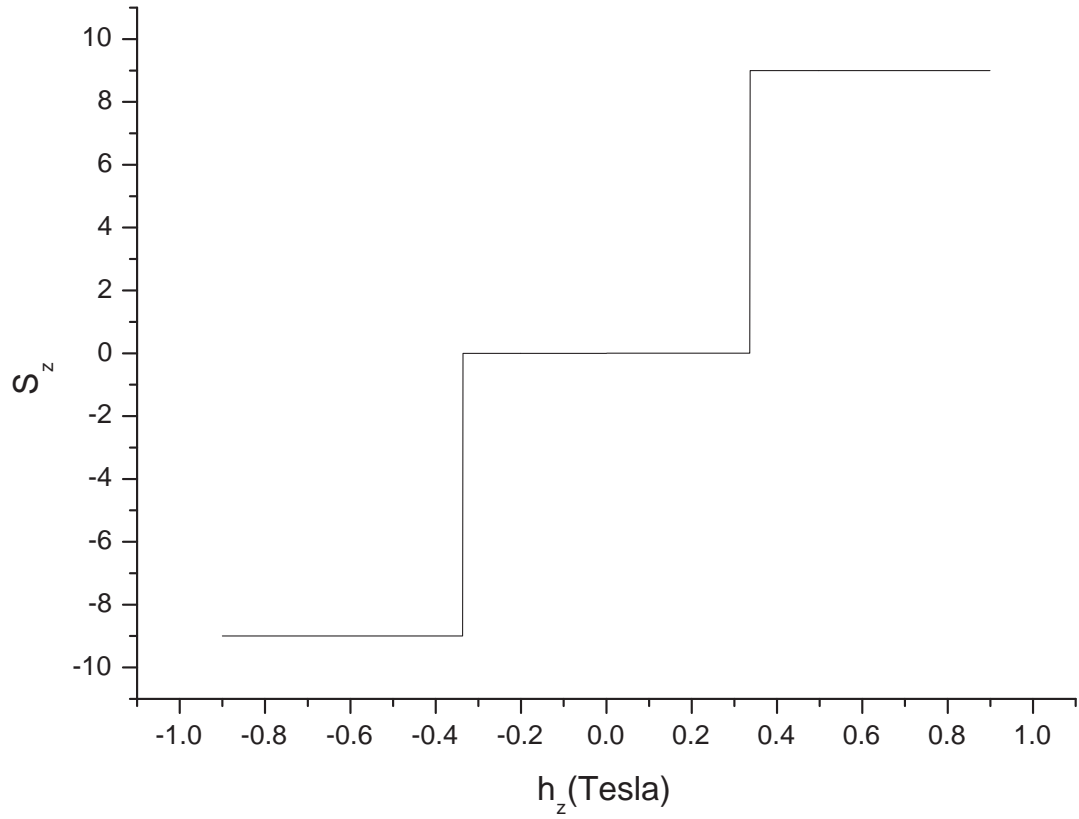}}
\caption{Magnetization curve based on a utmost-slow process. It
suppose that the state $\left| \Psi (t)\right\rangle $\ of system
always relax to the ground state $\left| \Phi (E_{\min
})\right\rangle $\ of $H(t)$.} \label{fig:epsart}
\end{figure}

\begin{figure}[tbp]
\center{\includegraphics[width=16cm]{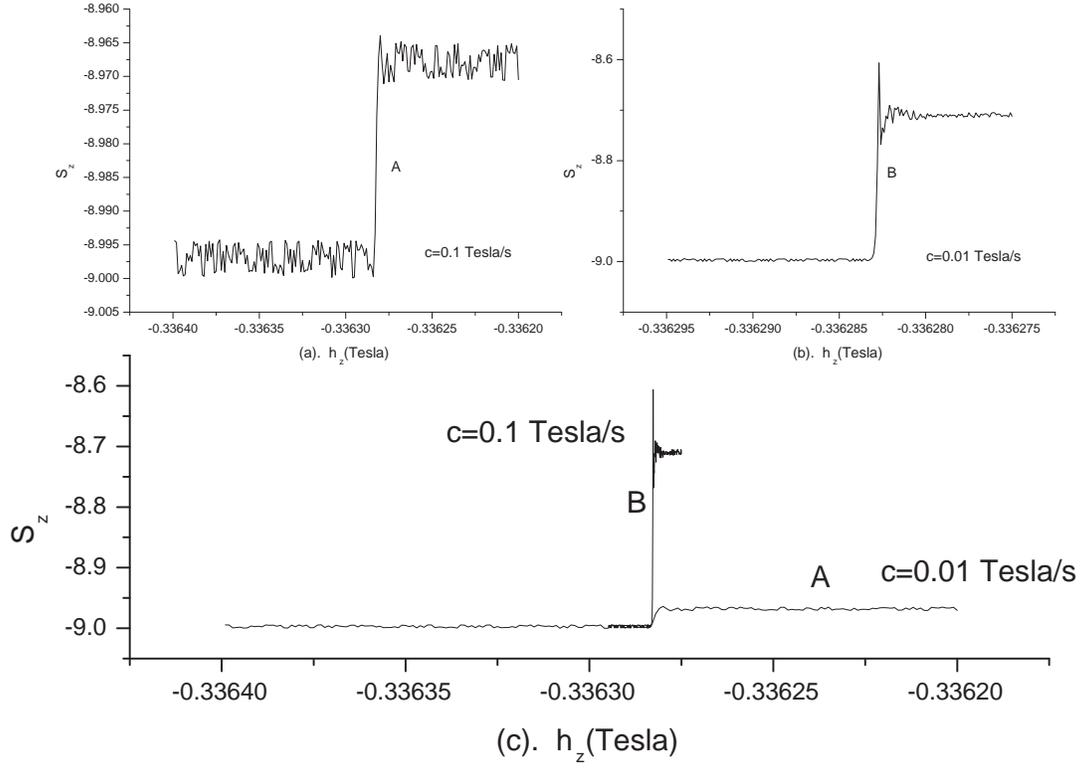}}
\caption{Magnetization curves response to slow sweeping fields
over very sharp ranges of $h_{z}$. The sweeping rates are the same
order with that used in experiment (Ref.$^{[11]}$).}
\label{fig:epsart}
\end{figure}

\end{document}